\documentclass[12pt,prd,showpacs,tightenlines,nofootinbib]{revtex4}
\usepackage{bm}
\usepackage{graphics}
\usepackage{rotating}
\usepackage{epsfig}
\begin{document}
\begin{flushright}{HU-EP-10/73}\end{flushright}
\title{Masses of tetraquarks with open charm and bottom}
\author{D. Ebert$^{1}$, R. N. Faustov$^{1,2}$  and V. O. Galkin$^{1,2}$}
\affiliation{
$^1$ Institut f\"ur Physik, Humboldt--Universit\"at zu Berlin,
Newtonstr. 15, D-12489  Berlin, Germany\\
$^2$ Dorodnicyn Computing Centre, Russian Academy of Sciences,
  Vavilov Str. 40, 119991 Moscow, Russia}

\begin{abstract}
 The masses of the heavy tetraquarks with open
charm and bottom  are calculated within  the
diquark-antidiquark picture in the framework of the relativistic quark
model. The dynamics of the light quarks and
diquarks is treated completely relativistically. The
diquark structure is taken into account by calculating
the diquark-gluon form factor. New experimental data on charmed and
charmed-strange mesons are discussed. Our results indicate that the
anomalous scalar $D_{s0}^*(2317)$ and axial vector $D_{s1}(2460)$
mesons could not be considered as 
diquark-antidiquark bound states. On the other hand, $D_s(2632)$ and
$D_{sJ}^*(2860)$ could be interpreted as scalar and tensor
tetraquarks, respectively. The predictions for masses of the corresponding bottom
counterparts of the charmed tetraquarks are given.   
\end{abstract}


\maketitle

\section{Introduction}
\label{sec:intr}

In last few years a significant experimental progress has been achieved in
charmonium and charmed meson spectroscopy. Many  new 
states, such as $X(3872)$, $Y(4260)$, $Y(4360)$, $Y(4660)$, $Z(4248)$,
$Z(4430)$, $D^*_{s0}(2317)$, $D_{s1}(2460)$, $D_s(2632)$ and
$D_{sJ}^*(2860)$ etc., were 
observed \cite{pakhlova} which cannot be simply accommodated in the 
quark-antiquark ($c\bar c$ and $c\bar s$) picture. These  states and especially the
charged charmonium-like ones
can be considered as indications of the possible existence of
exotic multiquark states. 

The open charm mesons, both with and without open strangeness,
represent a special interest. Even seven years after the  
discovery of the charmed-strange $D^*_{s0}(2317)$ and $D_{s1}(2460)$ mesons their
nature remains controversial in the literature. The abnormally light
masses of these mesons put them below $DK$ and $D^*K$ thresholds, thus
making these states narrow, since the only allowed strong  decays
$D^{(*)}_{sJ}\to D_s^{(*)}\pi$ violate isospin symmetry. The peculiar
feature of these mesons is that they 
have masses almost equal or even lower than the masses of their
charmed counterparts  $D^*_0(2400)$ and $D_1(2427)$ \cite{dexp,dsexp,pdg}. Most of the
theoretical approaches including lattice QCD \cite{lattice}, QCD sum
rule \cite{sr} and
different quark model \cite{egf,qm} calculations give masses of the $0^+$
and $1^+$ $P$-wave $c\bar s$ states significantly heavier (by 100-200 MeV) than
the measured ones. Different theoretical solutions of this problem
were proposed including consideration of these mesons as 
chiral partners of $0^-$ and $1^-$ states \cite{chiral}, $c\bar s$ states which are
strongly influenced by the nearby $DK$ thresholds \cite{cs}, $DK$ or
$D_s\pi$ molecules \cite{mol},
a mixture of $c\bar s$ and tetraquark states \cite{tetr,mppr,vfv,cgnps,gk}. However the universal
understanding of their nature is still missing. Therefore it is
very important to observe their bottom counterparts. The unquenched lattice
calculations of their masses can be found in Ref.~\cite{tmc}.

Another unexpectedly narrow charmed-strange meson $D_s(2632)$ was
discovered by SELEX Collaboration 
\cite{ds2632}. Its unusual decay properties triggered speculations
about its possible exotic origin \cite{pakhlova}. However, the status of
this state remains controversial since 
FOCUS \cite{focus2632}, BaBar \cite{babar2632} and Belle
\cite{belle2632} reported negative results in their search for this state.     

Three other charmed-strange mesons $D_{s1}(2710)$,
$D_{sJ}^*(2860)$ and $D_{sJ}(3040)$  were discovered at $B$-factories
by Belle and BaBar \cite{ds2710,ds3040}. Not all of them could be simply
accommodated in the usual $c\bar s$ picture. Their decay pattern
implies that $D_{sJ}^*(2860)$ should have natural parity, while
$D_{s1}(2710)$ and $D_{sJ}(3040)$ should have unnatural parity. Our
recent calculation of 
the heavy-light quark-antiquark meson spectra \cite{hlms} have shown
that $D_{s1}(2710)$ and $D_{sJ}(3040)$ are good candidates for the
$2^3S_1$ and  $2P_1$ states, respectively. They nicely fit to the
corresponding Regge trajectories, while $D^*_{s0}(2317)$, $D_{s1}(2460)$,
$D_{sJ}^*(2860)$ and $D_s(2632)$ have anomalously low masses (lower
than expected by 100-200 MeV) and do not lie on the respective Regge trajectories.

\begin{table}
\caption{Predictions \cite{hlms} for the masses of  charmed ($q=u,d$) and charmed-strange mesons
   (in MeV) in comparison with available experimental data \cite{pdg,babar2010}.} 
   \label{tab:csmm}
\begin{ruledtabular}
\begin{tabular}{cccccccc}
\multicolumn{2}{l}{\underline{\phantom{p}\hspace{0.8cm}State\hspace{0.8cm}}}&Theory
&\multicolumn{2}{c}{\underline{\hspace{1.6cm}Experiment\hspace{1.6cm}}}&
Theory&\multicolumn{2}{r}{\underline{\hspace{1.2cm}Experiment\hspace{1.2cm}}}\\
$n^{2S+1}L_J$&$J^{P}$&mass ($c\bar q$) & meson &mass&mass ($c\bar
s$) & meson&mass\\[2pt]
\hline
$1^1S_0$& $0^{-}$&1871& $D$&1869.62(16)&1969& $D_s$ & 1968.47(33)\\
$1^3S_1$& $1^{-}$&2010& $D^*(2010)$&2010.25(14)
&2111&$D^*_s$& 2112.3(5)\\
$1^3P_0$& $0^{+}$&2406&
$D_0^*(2400)$&$\left\{\begin{array}{l}2403(40)(^\pm)\\2318(29)(^0) \end{array}\right.$ &2509&$D_{s0}^*(2317)$&2317.8(6)\\
$1P_1$& $1^{+}$&2469& $D_1(2430)$&2427(40) &2574&$D_{s1}(2460)$&2459.5(6)\\
$1P_1$& $1^{+}$&2426& $D_1(2420)$&2423.4(3.1)&2536&$D_{s1}(2536)$&2535.29(20)\\
$1^3P_2$& $2^{+}$&2460& $D^*_2(2460)$&2460.1($^{+2.6}_{-3.5}$)&2571&$D_{s2}(2573)$&2572.6(9)\\

$2^1S_0$& $0^{-}$&2581& $D(2550)^\dag$ & 2539.4(8.1)$(^0)$ &2688&&\\
$2^3S_1$& $1^{-}$&2632& $\left\{\begin{array}{l}\\[-2ex]D^*(2600)^\dag\\[2ex] 
D^*(2640) \end{array}\right.$&$\begin{array}{c}\left\{\begin{array}{l}2621.3(5.6)(^+)\\
2608.7(3.5)(^0)\end{array}\right.\\  2637(6)? \end{array}$ &2731&$D_{s1}(2710)$& 2709($^{+9}_{-6}$)\\
$1^3D_1$& $1^{-}$&2788&$\left\{\begin{array}{l}\\[-2ex]D^*(2760)^\dag\\[2ex]D(2750)^\dag \end{array}\right.$
&$\begin{array}{c}\left\{\begin{array}{l}2769.7(4.1)(^+)\\
2763.3(3.3)(^0)\end{array}\right.\\2752.4(3.2) \end{array}$ &2913&&\\
$1D_2$& $2^{-}$&2850& & &2961&&\\
$1D_2$& $2^{-}$&2806& & &2931&&\\
$1^3D_3$& $3^{-}$&2863& & &2971&$D_{sJ}^*(2860)$& 2862($^{+6}_{-3}$)\\
$2^3P_0$& $0^{+}$&2919& & &3054&&\\
$2P_1$& $1^{+}$&3021& & &3154&&\\
$2P_1$& $1^{+}$&2932& & &3067&$D_{sJ}(3040)$ & 3044($^{+31}_{-9}$)\\
$2^3P_2$& $2^{+}$&3012& & &3142&&\\

$3^1S_0$& $0^{-}$&3062& & &3219&&\\
$3^3S_1$& $1^{-}$&3096& & &3242&& 
\end{tabular}
\end{ruledtabular}
$^\dag$ new states recently observed by BaBar \cite{babar2010}
\end{table}

In the charmed sector very recently the BaBar Collaboration discovered
four new signal peaks $D(2550)$, $D^*(2600)$, $D(2750)$ and
$D^*(2760)$ \cite{babar2010}. The
last two signals, observed in $D^*\pi$ and $D\pi$ modes,  have close
mass and width values (which differ by 2.6$\sigma$ and
1.5$\sigma$, respectively) and therefore could belong to the same state. 
The angular analysis shows that these
signals can be considered as candidates for the radially excited
$2^1S_0$,  $2^3S_1$  states and orbitally excited $1^3D_1$ state,
respectively. Their mass values are in a good 
agreement with the results of our model. We summarize our predictions
for the masses of the $c\bar q$ ($q=u,d$) and $c\bar s$ mesons in
Table~\ref{tab:csmm} and confront them with the available experimental
data \cite{pdg,babar2010}. As it is clearly seen there are 
indications of the possible existence 
of the exotic  states especially in the charmed-strange sector.

In  our  papers
\cite{tetr1,tetr2} we calculated masses of the
hidden  heavy-flavour tetraquarks and light tetraquarks in  the
framework of the relativistic quark 
model based on the quasipotential approach in quantum chromodynamics.
Here we extend this analysis to the consideration of 
heavy tetraquark states with open charm and bottom. This study could
help in revealing the nature of the anomalous charmed-strange mesons.

As previously \cite{tetr1,tetr2}, we use the  diquark-antidiquark
picture to reduce a complicated relativistic 
four-body problem to two subsequent more simple two-body
problems. The first step consists in the calculation of the masses, wave
functions and form factors of the diquarks, composed from light and heavy
quarks. At the second step, a tetraquark is considered to be a
bound diquark-antidiquark system. It is 
important to emphasize that we do not consider the diquark as a point
particle but explicitly take into account its structure by calculating
the form factor of the diquark-gluon interaction in terms of the
diquark wave functions.

\section{Relativistic model of tetraquarks}
\label{sec:rqm}

In the quasipotential approach and diquark-antidiquark picture of
heavy tetraquarks the interaction of two quarks in a diquark and
the diquark-antidiquark interaction in a tetraquark are described
by the diquark wave function ($\Psi_{d}$) of the bound quark-quark
state and by the tetraquark wave function ($\Psi_{T}$) of the
bound diquark-antidiquark state, respectively. These wave functions satisfy the
quasipotential equation of the Schr\"odinger type \cite{efg}
\begin{equation}
\label{quas}
{\left(\frac{b^2(M)}{2\mu_{R}}-\frac{{\bf
p}^2}{2\mu_{R}}\right)\Psi_{d,T}({\bf p})} =\int\frac{d^3 q}{(2\pi)^3}
 V({\bf p,q};M)\Psi_{d,T}({\bf q}),
\end{equation}
where the relativistic reduced mass is
\begin{equation}
\mu_{R}=\frac{E_1E_2}{E_1+E_2}=\frac{M^4-(m^2_1-m^2_2)^2}{4M^3},
\end{equation}
and $E_1$, $E_2$ are given by
\begin{equation}
\label{ee}
E_1=\frac{M^2-m_2^2+m_1^2}{2M}, \quad E_2=\frac{M^2-m_1^2+m_2^2}{2M}.
\end{equation}
Here, $M=E_1+E_2$ is the bound-state mass (diquark or tetraquark),
$m_{1,2}$ are the masses of quarks ($q$ and $Q$) which form
the diquark or of the diquark ($d$) and antiquark ($\bar d'$) which
form the heavy tetraquark ($T$), and ${\bf p}$ is their relative
momentum. In the center-of-mass system the relative momentum
squared on mass shell reads
\begin{equation}
{b^2(M) }
=\frac{[M^2-(m_1+m_2)^2][M^2-(m_1-m_2)^2]}{4M^2}.
\end{equation}

The kernel $V({\bf p,q};M)$ in Eq.~(\ref{quas}) is the
quasipotential operator of the quark-quark or diquark-antidiquark
interaction. It is constructed with the help of the off-mass-shell
scattering amplitude, projected onto the positive-energy states.
 For the quark-quark interaction in a diquark we
use the relation $V_{qq}=V_{q\bar q}/2$ arising under the
assumption of an octet structure of the interaction from the
difference in the $qq$ and $q\bar q$ colour
states. An important role in this construction is played by the
Lorentz structure of the confining interaction. In our analysis of
mesons, while constructing the quasipotential of the
quark-antiquark interaction, we assumed that the effective
interaction is the sum of the usual one-gluon exchange term and a
mixture of long-range vector and scalar linear confining
potentials, where the vector confining potential contains the
Pauli term. We use the same conventions for the construction of
the quark-quark and diquark-antidiquark interactions in the
tetraquark. The quasipotential is then defined as follows
\cite{egf,tetr1}.

(a) For the quark-quark  ($Qq$) interactions,
$V({\bf p,q};M)$ reads
 \begin{equation}
\label{qpot}
V({\bf p,q};M)=\bar{u}_{1}(p)\bar{u}_{2}(-p){\cal V}({\bf p}, {\bf
q};M)u_{1}(q)u_{2}(-q),
\end{equation}
with
\[
{\cal V}({\bf p,q};M)=\frac12\left[\frac43\alpha_sD_{ \mu\nu}({\bf
k})\gamma_1^{\mu}\gamma_2^{\nu}+ V^V_{\rm conf}({\bf k})
\Gamma_1^{\mu}({\bf k})\Gamma_{2;\mu}(-{\bf k})+
 V^S_{\rm conf}({\bf k})\right].
\]
Here, $\alpha_s$ is the QCD coupling constant; $D_{\mu\nu}$ is the
gluon propagator in the Coulomb gauge,
\begin{equation}
D^{00}({\bf k})=-\frac{4\pi}{{\bf k}^2}, \quad D^{ij}({\bf k})=
-\frac{4\pi}{k^2}\left(\delta^{ij}-\frac{k^ik^j}{{\bf k}^2}\right),
\quad D^{0i}=D^{i0}=0,
\end{equation}
and ${\bf k=p-q}$; $\gamma_{\mu}$ and $u(p)$ are the Dirac
matrices and spinors,
\begin{equation}
\label{spinor}
u^\lambda({p})=\sqrt{\frac{\epsilon(p)+m}{2\epsilon(p)}}
\left(\begin{array}{c} 1\\
\displaystyle\frac{\mathstrut\bm{\sigma}\cdot{\bf p}}
{\mathstrut\epsilon(p)+m}
\end{array}\right)
\chi^\lambda,
\end{equation}
with $\epsilon(p)=\sqrt{{\bf p}^2+m^2}$.

The effective long-range vector vertex of the quark is
defined by \cite{egf}
\begin{equation}
\Gamma_{\mu}({\bf k})=\gamma_{\mu}+
\frac{i\kappa}{2m}\sigma_{\mu\nu}\tilde k^{\nu}, \qquad \tilde
k=(0,{\bf k}),
\end{equation}
where $\kappa$ is the Pauli interaction constant characterizing
the anomalous chromomagnetic moment of quarks. In configuration
space the vector and scalar confining potentials in the
nonrelativistic limit \cite{efg} reduce to
\begin{eqnarray}
V^V_{\rm conf}(r)&=&(1-\varepsilon)V_{\rm
conf}(r),\nonumber\\[1ex] V^S_{\rm conf}(r)& =&\varepsilon V_{\rm
conf}(r),
\end{eqnarray}
with
\begin{equation}
V_{\rm conf}(r)=V^S_{\rm conf}(r)+
V^V_{\rm conf}(r)=Ar+B,
\end{equation}
where $\varepsilon$ is the mixing coefficient.

(b) For the diquark-antidiquark ($d\bar d'$) interaction, $V({\bf
p,q};M)$ is given by
\begin{eqnarray}
\label{dpot} V({\bf p,q};M)&=&\frac{\langle
d(P)|J_{\mu}|d(Q)\rangle} {2\sqrt{E_dE_d}} \frac43\alpha_sD^{
\mu\nu}({\bf k})\frac{\langle d'(P')|J_{\nu}|d'(Q')\rangle}
{2\sqrt{E_{d'}E_{d'}}}\nonumber\\[1ex]
&&+\psi^*_d(P)\psi^*_{d'}(P')\left[J_{d;\mu}J_{d'}^{\mu} V_{\rm
conf}^V({\bf k})+V^S_{\rm conf}({\bf
k})\right]\psi_d(Q)\psi_{d'}(Q'),
\end{eqnarray}
where $\langle
d(P)|J_{\mu}|d(Q)\rangle$ is the vertex of the
diquark-gluon interaction which takes into account the finite size of
the diquark
$\Big[$$P^{(')}=(E_{d^{(')}},\pm{\bf p})$ and
$Q^{(')}=(E_{d^{(')}},\pm{\bf q})$,
$E_d=(M^2-M_{d'}^2+M_d^2)/(2M)$ and $E_{d'}=(M^2-M_d^2+M_{d'}^2)/(2M)$
$\Big]$.

The diquark state in the confining part of the diquark-antidiquark
quasipotential (\ref{dpot}) is described by the wave functions
\begin{equation}
 \label{eq:ps}
 \psi_d(p)=\left\{\begin{array}{ll}1 &\qquad \text{for a scalar
 diquark,}\\[1ex]
\varepsilon_d(p) &\qquad \text{for an axial-vector diquark,}
\end{array}\right.
\end{equation}
where the four-vector
\begin{equation}\label{pv}
\varepsilon_d(p)=\left(\frac{(\bm{\varepsilon}_d\cdot{\bf
p})}{M_d},\bm{\varepsilon}_d+ \frac{(\bm{\varepsilon}_d\cdot{\bf
p}){\bf
 p}}{M_d(E_d(p)+M_d)}\right), \qquad \varepsilon^\mu_d(p) p_\mu=0,
\end{equation}
is the polarization vector of the axial-vector diquark with
momentum ${\bf p}$, $E_d(p)=\sqrt{{\bf p}^2+M_d^2}$, and
$\varepsilon_d(0)=(0,\bm{\varepsilon}_d)$ is the polarization
vector in the diquark rest frame. The effective long-range vector
vertex of the diquark can be presented in the form
\begin{equation}
 \label{eq:jc}
 J_{d;\mu}=\left\{\begin{array}{ll}
 \frac{\displaystyle (P+Q)_\mu}{\displaystyle
 2\sqrt{E_dE_d}}&\qquad \text{ for a scalar diquark,}\\[3ex]
-\; \frac{\displaystyle (P+Q)_\mu}{\displaystyle2\sqrt{E_dE_d}}
 +\frac{\displaystyle i\mu_d}{\displaystyle 2M_d}\Sigma_\mu^\nu
\tilde k_\nu
 &\qquad \text{ for an axial-vector diquark.}\end{array}\right.
\end{equation}
Here, the antisymmetric tensor
$\Sigma_\mu^\nu$ is defined by
\begin{equation}
 \label{eq:Sig}
 \left(\Sigma_{\rho\sigma}\right)_\mu^\nu=-i(g_{\mu\rho}\delta^\nu_\sigma
 -g_{\mu\sigma}\delta^\nu_\rho),
\end{equation}
and the axial-vector diquark spin ${\bf S}_d$ is given by
$(S_{d;k})_{il}=-i\varepsilon_{kil}$; $\mu_d$ is the total
chromomagnetic moment of the axial-vector diquark.

The constituent quark masses $m_c=1.55$ GeV, $m_b=4.88$ GeV,
$m_u=m_d=0.33$ GeV, $m_s=0.5$ GeV and the parameters of the linear
potential $A=0.18$ GeV$^2$ and $B=-0.3$~GeV have values
typical in quark models. The value of the mixing coefficient of
vector and scalar confining potentials $\varepsilon=-1$ has been
determined from the consideration of charmonium radiative decays
\cite{efg} and the heavy-quark expansion \cite{fg}. The universal
Pauli interaction constant $\kappa=-1$ has been fixed from the
analysis of the fine splitting of heavy quarkonia ${ }^3P_J$ -
states \cite{efg}. In this case, the long-range chromomagnetic
interaction of quarks vanishes in accordance with the flux-tube
model.

Since we deal with diquarks and tetraquarks containing light quarks
and diquarks, respectively, we adopt for the QCD
coupling constant $\alpha_s(\mu^2)$ the
simplest model with freezing \cite{bvb}, namely
\begin{equation}
  \label{eq:alpha}
  \alpha_s(\mu^2)=\frac{4\pi}{\displaystyle\beta_0
\ln\frac{\mu^2+M_B^2}{\Lambda^2}}, \qquad \beta_0=11-\frac23n_f,
\end{equation}
where the scale is taken as $\mu=2m_1
m_2/(m_1+m_2)$, the background mass is $M_B=2.24\sqrt{A}=0.95$~GeV \cite{bvb}, and
the parameter $\Lambda=413$~MeV was fixed from fitting the $\rho$
mass \cite{lregge}. Note that the other popular
parametrization of $\alpha_s$ with freezing \cite{shirkov} leads to close
values.

\section{Diquark and tetraquark masses}
\label{sec:dtm}

At the first step, we calculate the masses and form factors of the
 heavy and light
diquarks. Since the light quarks are highly relativistic
a completely relativistic treatment of the light quark dynamics is
required. To achieve this goal,  we closely follow our consideration
of diquarks in heavy baryons
and adopt the same procedure to make the relativistic
potential local by replacing
$\epsilon_{1,2}(p)=\sqrt{m_{1,2}^2+{\bf p}^2}\to E_{1,2}=(M^2-m_{2,1}^2+m_{1,2}^2)/2M$. 
Solving numerically the quasipotential equation (\ref{quas}) with the
complete relativistic potential,  which depends on the
diquark mass in a complicated highly nonlinear way \cite{hbar}, we get
the diquark masses and wave functions. In order to determine the
diquark interaction with the gluon field, which 
takes into account the diquark structure, we
calculate the corresponding matrix element of the quark
current between diquark states. Such calculation leads to the
emergence of the form factor $F(r)$ entering the vertex of the
diquark-gluon interaction \cite{hbar}. This form factor is expressed
through the overlap integral of the diquark wave functions. Our estimates show that this form factor can be approximated  with a
high accuracy by the expression 
\begin{equation}
  \label{eq:fr}
  F(r)=1-e^{-\xi r -\zeta r^2}.
\end{equation}
The values of the masses and parameters $\xi$ and $\zeta$ for light and heavy
scalar diquark $[\cdots]$ and axial vector diquark $\{\cdots\}$ ground
states were calculated previously \cite{hbar,tetr1} and are
given in Table~\ref{tab:dmass}.

\begin{table}
  \caption{Masses $M$ and form factor  parameters of
    diquarks. $S$ and $A$ 
    denote scalar and axial vector diquarks which are antisymmetric $[\cdots]$ and
    symmetric $\{\cdots\}$ in flavour, respectively. }
  \label{tab:dmass}
\begin{ruledtabular}
\begin{tabular}{ccccc}
Quark& Diquark&   $M$ &$\xi$ & $\zeta$
 \\
content &type & (MeV)& (GeV)& (GeV$^2$)  \\
\hline
$[u,d]$&S & 710 & 1.09 & 0.185  \\
$\{u,d\}$&A & 909 &1.185 & 0.365  \\
$[u,s]$& S & 948 & 1.23 & 0.225 \\
$\{u,s\}$& A & 1069 & 1.15 & 0.325\\
$\{s,s\}$ & A& 1203 & 1.13 & 0.280\\
$[c,q]$& $S$ & 1973& 2.55 &0.63  \\
$\{c,q\}$& $A$ & 2036& 2.51 &0.45  \\
$[c,s]$ & $S$& 2091& 2.15 & 1.05  \\
$\{c,s\}$& $A$ & 2158&2.12& 0.99 \\
$[b,q]$& $S$ & 5359 &6.10 & 0.55 \\
$\{b,q\}$& $A$ & 5381& 6.05 &0.35 \\
$[b,s]$ & $S$& 5462 & 5.70 &0.35 \\
$\{b,s\}$& $A$ & 5482 & 5.65 &0.27
  \end{tabular}
\end{ruledtabular}
\end{table}

At the second step, we calculate the masses of heavy tetraquarks 
considered as the bound states of a heavy-light diquark and light
antidiquark. For the
potential of the $S$-wave ($\langle{\bf L}^2\rangle=0$) diquark-antidiquark interaction (\ref{dpot}) we
get \cite{tetr2}  
\begin{eqnarray}
 \label{eq:pot}
 V(r)&=& \hat V_{\rm Coul}(r)+V_{\rm conf}(r)
+\frac1{E_1E_2}\Biggl\{{\bf
 p}\left[\hat V_{\rm Coul}(r)+V^V_{\rm conf}(r)\right]{\bf p} -\frac14
\Delta V^V_{\rm conf}(r)\nonumber\\[1ex]
&&
+\frac23\left[\Delta \hat V_{\rm
Coul}(r)+\frac{\mu_d^2}4\frac{E_1E_2}{M_1M_2} \Delta V^V_{\rm
conf}(r)\right]{\bf S}_1\cdot{\bf S}_2\Biggr\},
\end{eqnarray}
where $$\hat V_{\rm Coul}(r)=-\frac{4}{3}\alpha_s
\frac{F_1(r)F_2(r)}{r}$$ is the Coulomb-like one-gluon exchange
potential which takes into account the finite sizes of the diquark
and antidiquark through corresponding form factors $F_{1,2}(r)$.
Here, ${\bf S}_{1,2}$ are the spin operators of
diquark and antidiquark.  In the
following we choose the total chromomagnetic moment of the
axial-vector diquark $\mu_d=0$. Such a choice appears to be
natural, since the long-range chromomagnetic interaction of
diquarks proportional to $\mu_d$ then also vanishes in accordance
with the flux-tube model.

The resulting quasipotential equation with the complete kernel
(\ref{eq:pot}) is solved numerically without any approximations.

\section{Results and discussion}
\label{sec:rd}

\begin{table}
  \caption{Masses of charmed and bottom diquark-antidiquark ground ($1S$) states
    (in MeV) and possible experimental candidates \cite{pdg}. $S$ and $A$
    denote scalar and axial vector diquarks. }
  \label{tab:tmass}
\begin{ruledtabular}
\begin{tabular}{cccccc}
State& Diquark & Theory&
\multicolumn{2}{l}{\underline{\hspace{2.2cm}Experiment\hspace{2.2cm}}} 
& Theory \\
$J^{P}$ & content& Mass& Meson  & Mass & Mass \\
\hline\hline\\[-2.5ex]
&&$\bm{cq\bar q\bar q}$&&&$\bm{bq\bar q\bar q}$\\[1ex]
$0^{+}$ & $S\bar S$ & 2399& $D_0^*(2400)$ &$\left\{\begin{array}{l}2403(40)(^\pm)\\2318(29)(^0) \end{array}\right.$ & 5758\\
$1^{+}$ & $S\bar A$& 2558&  & &5950\\
$1^{+}$ & $A\bar S$& 2473& $D_1(2430)$&2427(40) &5782\\
$0^{+}$& $A\bar A$ & 2503 & & &5896\\
$1^{+}$& $A\bar A$ & 2580 & & &5937\\
$2^{+}$& $A\bar A$ & 2698 & & &6007\\[1.ex]\hline\\[-2ex]
&&$\bm{cq\bar s\bar q}$&&&$\bm{bq\bar s\bar q$}\\[1ex]
$0^{+}$ & $S\bar S$ & 2619& $D_s(2632)$ &2632.5(1.7) & 5997\\
$1^{+}$ & $S\bar A$& 2723&  & &6125\\
$1^{+}$ & $A\bar S$& 2678& & &6021\\
$0^{+}$& $A\bar A$ & 2689 & & &6086\\
$1^{+}$& $A\bar A$ & 2757 & & &6118\\
$2^{+}$& $A\bar A$ & 2863&$D^*_{sJ}(2860)$ 
&$2862\left(^{+6}_{-3}\right)$ &6177\\[1.ex]\hline\\[-2ex]
&&$\bm{cs\bar s\bar q}$&&&$\bm{bs\bar s\bar q}$\\[1ex]
$0^{+}$ & $S\bar S$ & 2753 & & & 6108\\
$1^{+}$ & $S\bar A$& 2870&  & &6238\\
$1^{+}$ & $A\bar S$& 2830& & &6134\\
$0^{+}$& $A\bar A$ & 2839 & & &6197\\
$1^{+}$& $A\bar A$ & 2901 & & &6228\\
$2^{+}$& $A\bar A$ & 2998 &&&6284\\[1.ex]\hline\\[-2ex]
&&$\bm{cs\bar s\bar s}$&&&$\bm{bs\bar s\bar s}$\\[1ex]
$1^{+}$ & $S\bar A$& 3025&  & &6383\\
$0^{+}$& $A\bar A$ & 3003 & & &6353\\
$1^{+}$& $A\bar A$ & 3051 & & &6372\\
$2^{+}$& $A\bar A$ & 3135 &&&6411
 \end{tabular}
\end{ruledtabular}
\end{table}

Masses of the heavy tetraquark ground ($1S$) states calculated in
the diquark-antidiquark picture are presented in
Table~\ref{tab:tmass}. In this table we also give possible
experimental candidates for charmed and charmed-strange
tetraquarks. 

In the charmed meson sector the situation is rather
complicated. Comparing results presented in Tables~\ref{tab:csmm} and
\ref{tab:tmass} we see that our model predicts very close masses for the
scalar $0^+$ orbitally excited $c\bar q$ meson  ($1^3P_0$) and the ground
state ($1S$) tetraquark, composed from the scalar $[cq]$ diquark and
scalar $[\bar q\bar q]$
antidiquark. The same is true for the masses of the $1^+$ axial vector $c\bar q$
 meson ($1P_1$) and the tetraquark, composed from the axial vector $\{cq\}$
diquark and scalar $[\bar q\bar q]$ antidiquark. The calculated masses are
consistent with the measured masses of the scalar $D_0^*(2400)$ and axial
vector $D_1(2430)$ mesons. The mixing of the 
$c\bar q$ and tetraquark states could be responsible for the 
observed difference in masses of the charged and neutral $D_0^*(2400)$
mesons.

In the charmed-strange sector  the  $0^+$ and $1^+$
tetraquarks are predicted to have masses significantly (by 200-300 MeV) higher than 
experimentally measured masses of the $D^*_{s0}(2317)$ and
$D_{s1}(2460)$ mesons (cf. Tables~\ref{tab:csmm},
\ref{tab:tmass}). This excludes the interpretation of these anomalously
light $D_s$ mesons as the heavy diquark-antidiquark (tetraquark)
states in our model. Instead, we find that the lightest scalar $0^+$ 
tetraquark, composed from the scalar $[cq]$ diquark and
scalar $[\bar s\bar q]$ antidiquark, has a mass consistent with the
controversial  $D_s(2632)$  observed by SELEX \cite{ds2632}. The
$D_{sJ}^*(2860)$ meson, observed by BaBar \cite{ds3040} both in
$DK$ and $D^*K$ modes,\footnote{This state should therefore have
  natural parity and total spin $J\ge 1$.} has mass coinciding within
experimental error bars with the prediction  for the mass of the
tensor $2^+$ tetraquark,  composed form the axial vector 
$\{cq\}$ diquark and axial vector $\{\bar s\bar q\}$ antidiquark. The
rather high value of its spin can explain the non-observation of this
state by Belle in $B$ decays.   

\begin{table}
  \caption{Comparison of theoretical predictions for the masses of
    $cq\bar s\bar q$ tetraquarks (in MeV).}
  \label{tab:ecmass}
\begin{ruledtabular}
\begin{tabular}{cccccc}
$J^{P}$ & this paper & \cite{mppr} & \cite{cgnps} & \cite{vfv} & \cite{gk}\\
\hline
$0^{+}$ &2619& 2371 & 2840 & 2731 & 2616\\
$1^{+}$ & 2678 & 2410 & 2841&\\
$1^{+}$ & 2723 & 2462 & 2880 & 2841& \\
$0^{+}$& 2689 & 2424 &2503 & 2699&\\
$1^{+}$& 2757& 2571& 2748&\\
$2^{+}$& 2863& 2648 & 2983& & 2854\\
 \end{tabular}
\end{ruledtabular}
\end{table}

In Table~\ref{tab:ecmass} we compare our results  for
the masses of the charmed-strange 
diquark-antidiquark bound states with the predictions of
Refs.~\cite{mppr,cgnps,vfv,gk}. From this table we see that only the
model of Ref.~\cite{mppr} gives masses of scalar and axial vector
tetraquarks compatible with the observed masses of the $D^*_{s0}(2317)$ and
$D_{s1}(2460)$ mesons. All other models predict significantly higher
mass values for these states. The main difference between these
approaches consists in the substantial distinctions in treating quark
dynamics in tetraquarks. The authors of Ref.~\cite{mppr} use a
phenomenological approach, determining diquark masses and parameters
of hyperfine interactions between quarks from adjusting their
predictions to experimental observables. Contrary we describe diquarks
and tetraquarks dynamically as quark-quark and diquark-antidiquark
bound systems and calculate their masses and form factors in the model
where all parameters were previously fixed from considerations of
meson properties. Different dynamical approaches were applied in
Refs.~\cite{cgnps,vfv,gk}. The authors of Ref.~\cite{cgnps} calculate
diquark-antidiquark mass spectra in the quark model employing the QCD potential
found by means of the AdS/QCD correspondence. Tetraquark masses are
calculated in Ref.~\cite{vfv} in the nonrelativistic quark model
including both the confining interaction and meson exchanges, while in
Ref.\cite{gk} the coupled-channel formalism is employed. In the latter two
approaches the anomalous charmed-strange mesons could be only
accommodated as a mixture of quark-antiquark and tetraquark states with
a phenomenologically adjusted mixing interaction.   Thus it seems to be 
unlikely that the $D^*_{s0}(2317)$ and $D_{s1}(2460)$ could be pure
diquark-antidiquark bound states.       

\section{Conclusions}
\label{sec:conc}

We calculated the masses of heavy tetraquarks with open
charm and bottom in the diquark-antidiquark picture using the dynamical approach
based on the relativistic quark model. Both diquark and tetraquark
masses were obtained by the numerical solution of the quasipotential wave
equation with the corresponding relativistic 
potentials. The diquark structure was taken into account in terms of
diquark wave functions. It is important to emphasize  
that, in our analysis, we did not introduce any free adjustable
parameters but used their values fixed from our previous considerations
of heavy and light hadron properties. It was found that the
$D_0^*(2400)$, $D_s(2632)$ and $D_{sJ}^*(2860)$   mesons 
could be tetraquark states with open charm, while the $D^*_{s0}(2317)$
and $D_{s1}(2460)$ mesons cannot be interpreted as 
diquark-antidiquark bound states. The masses of the bottom counterparts of
charmed tetraquarks were calculated. It is important to search
for them in order to help revealing the nature of controversial
charmed and charmed-strange mesons.

The authors are grateful to M. M\"uller-Preussker  for support  and to
V. Matveev, V. Savrin and M. Wagner for discussions.  This work was supported in
part by  the Deutsche 
Forschungsgemeinschaft (DFG) under contract Eb 139/6-1 and the Russian
Foundation for Basic Research (RFBR) grants
No.08-02-00582 and No.10-02-91339.


\begin{thebibliography}{99}
\bibitem{pakhlova} For recent reviews see e.g. N.~Brambilla {\it et al.},
  arXiv:1010.5827 [hep-ph]; 
N.~Drenska, R.~Faccini, F.~Piccinini, A.~Polosa, F.~Renga and C.~Sabelli,
  arXiv:1006.2741 [hep-ph];
 S.~Godfrey and S.~L.~Olsen,
  Ann.\ Rev.\ Nucl.\ Part.\ Sci.\  {\bf 58}, 51 (2008); E. S. Swanson,
  Phys. Rept. {\bf 429}, 243 (2006). 
\bibitem{dsexp}
  B.~Aubert {\it et al.}  [BABAR Collaboration],
  Phys.\ Rev.\ Lett.\  {\bf 90}, 242001 (2003);
D.~Besson {\it et al.}  [CLEO Collaboration],
  Phys.\ Rev.\  D {\bf 68}, 032002 (2003)
  [Erratum-ibid.\  D {\bf 75}, 119908 (2007)];
K.~Abe {\it et al.},
  Phys.\ Rev.\ Lett.\  {\bf 92}, 012002 (2004);
B.~Aubert {\it et al.}  [BABAR Collaboration],
  Phys.\ Rev.\  D {\bf 69}, 031101 (2004).

\bibitem{dexp}
  K.~Abe {\it et al.}  [Belle Collaboration],
  Phys.\ Rev.\  D {\bf 69}, 112002 (2004);
J.~M.~Link {\it et al.}  [FOCUS Collaboration],
  Phys.\ Lett.\  B {\bf 586}, 11 (2004);
S.~Anderson {\it et al.}  [CLEO Collaboration],
  Nucl.\ Phys.\  A {\bf 663}, 647 (2000);
I.~V.~Gorelov  [CDF Collaboration],
  J.\ Phys.\ Conf.\ Ser.\  {\bf 9}, 67 (2005).
\bibitem{pdg}
  K.~Nakamura  [Particle Data Group],
  J.\ Phys.\ G {\bf 37}, 075021 (2010).

\bibitem{lattice}
  R.~Lewis and R.~M.~Woloshyn,
  Phys.\ Rev.\  D {\bf 62}, 114507 (2000);
G.~S.~Bali,
  Phys.\ Rev.\  D {\bf 68}, 071501 (2003);
A.~Dougall, R.~D.~Kenway, C.~M.~Maynard and C.~McNeile  [UKQCD
                  Collaboration],
  Phys.\ Lett.\  B {\bf 569}, 41 (2003);
D.~Mohler and R.~M.~Woloshyn,
  arXiv:1010.2786 [hep-lat].

\bibitem{sr}
Y.~B.~Dai, C.~S.~Huang, C.~Liu and S.~L.~Zhu,
  Phys.\ Rev.\  D {\bf 68}, 114011 (2003);
  S.~Narison,
  Phys.\ Lett.\  B {\bf 605}, 319 (2005).


\bibitem{egf} D.~Ebert, V.~O.~Galkin and R.~N.~Faustov,
  Phys.\ Rev.\  D {\bf 57}, 5663 (1998)
  [Erratum-ibid.\  D {\bf 59}, 019902 (1999)].

\bibitem{qm}
  S.~Godfrey and N.~Isgur,
  Phys.\ Rev.\  D {\bf 32}, 189 (1985);
M.~Di Pierro and E.~Eichten,
  Phys.\ Rev.\  D {\bf 64}, 114004 (2001)
Yu.~S.~Kalashnikova, A.~V.~Nefediev and Yu.~A.~Simonov,
  Phys.\ Rev.\  D {\bf 64}, 014037 (2001); 
S.~Godfrey,
  Phys.\ Rev.\  D {\bf 72}, 054029 (2005).
\bibitem{chiral}
  W.~A.~Bardeen, E.~J.~Eichten and C.~T.~Hill,
  Phys.\ Rev.\  D {\bf 68}, 054024 (2003); 
M.~A.~Nowak, M.~Rho and I.~Zahed,
  Phys.\ Rev.\  D {\bf 48}, 4370 (1993);
D.~Ebert, T.~Feldmann, R.~Friedrich and H.~Reinhardt,
  Nucl.\ Phys.\  B {\bf 434}, 619 (1995);
D.~Ebert, T.~Feldmann and H.~Reinhardt,
  Phys.\ Lett.\  B {\bf 388}, 154 (1996).

\bibitem{cs}
  E.~van Beveren and G.~Rupp,
  Phys.\ Rev.\ Lett.\  {\bf 91}, 012003 (2003);
D.~S.~Hwang and D.~W.~Kim,
  Phys.\ Lett.\  B {\bf 601}, 137 (2004);
A.~M.~Badalian, Yu.~A.~Simonov and M.~A.~Trusov,
  Phys.\ Rev.\  D {\bf 77}, 074017 (2008).


\bibitem{mol}
  T.~Barnes, F.~E.~Close and H.~J.~Lipkin,
  Phys.\ Rev.\  D {\bf 68}, 054006 (2003);
A.~P.~Szczepaniak,
  Phys.\ Lett.\  B {\bf 567}, 23 (2003);
A.~Faessler, T.~Gutsche, V.~E.~Lyubovitskij and Y.~L.~Ma,
  Phys.\ Rev.\  D {\bf 76}, 114008 (2007).

\bibitem{tetr}
  K.~Terasaki,
  Phys.\ Rev.\  D {\bf 68}, 011501 (2003);
M.~E.~Bracco, A.~Lozea, R.~D.~Matheus, F.~S.~Navarra and M.~Nielsen,
  Phys.\ Lett.\  B {\bf 624}, 217 (2005).


\bibitem{mppr}
L.~Maiani, F.~Piccinini, A.~D.~Polosa and V.~Riquer,
  Phys.\ Rev.\  D {\bf 71}, 014028 (2005).


\bibitem{cgnps}
M.~V.~Carlucci, F.~Giannuzzi, G.~Nardulli, M.~Pellicoro and S.~Stramaglia,
  Eur.\ Phys.\ J.\  C {\bf 57}, 569 (2008).

\bibitem{vfv}
J.~Vijande, F.~Fernandez and A.~Valcarce,
  Phys.\ Rev.\  D {\bf 73}, 034002 (2006)
  [Erratum-ibid.\  D {\bf 74}, 059903 (2006)].

\bibitem{gk}
  S.~M.~Gerasyuta and V.~I.~Kochkin,
  Phys.\ Rev.\  D {\bf 78}, 116004 (2008).


\bibitem{tmc}
  K.~Jansen, C.~Michael, A.~Shindler and M.~Wagner  [ETM Collaboration],
  JHEP {\bf 0812}, 058 (2008).

\bibitem{ds2632}
A.~V.~Evdokimov {\it et al.}  [SELEX Collaboration],
  Phys.\ Rev.\ Lett.\  {\bf 93}, 242001 (2004).


\bibitem{focus2632}
R. Kutschke [FOCUS Collaboration], E831-doc-701-v2 (2004).

\bibitem{babar2632}
  B.~Aubert {\it et al.}  [BABAR Collaboration],
  arXiv:hep-ex/0408087.
\bibitem{belle2632}
  B.~D.~Yabsley  [Belle Collaboration],
  AIP Conf.\ Proc.\  {\bf 792}, 875 (2005).

\bibitem{ds2710}
  B.~Aubert {\it et al.}  [BABAR Collaboration],
  Phys.\ Rev.\ Lett.\  {\bf 97}, 222001 (2006);
J.~Brodzicka {\it et al.}  [Belle Collaboration],
  Phys.\ Rev.\ Lett.\  {\bf 100}, 092001 (2008).

\bibitem{ds3040}
 B.~Aubert   {\it et al.}  [BABAR Collaboration],
  Phys.\ Rev.\  D {\bf 80}, 092003 (2009).

\bibitem{hlms}
  D.~Ebert, R.~N.~Faustov and V.~O.~Galkin,
  Eur.\ Phys.\ J.\  C {\bf 66}, 197 (2010).

\bibitem{babar2010}
  P.~del Amo Sanchez {\it et al.}  [The BABAR Collaboration],
  arXiv:1009.2076 [hep-ex].

\bibitem{tetr1}
  D.~Ebert, R.~N.~Faustov and V.~O.~Galkin,
  Phys.\ Lett.\  B {\bf 634}, 214 (2006); 
  Eur.\ Phys.\ J.\  C {\bf 58}, 399 (2008); 
  Mod.\ Phys.\ Lett.\  A {\bf 24}, 567 (2009); D.~Ebert, R.~N.~Faustov, V.~O.~Galkin and W.~Lucha,
  Phys.\ Rev.\  D {\bf 76}, 114015 (2007).

\bibitem{tetr2}
  D.~Ebert, R.~N.~Faustov and V.~O.~Galkin,
  Eur.\ Phys.\ J.\  C {\bf 60}, 273 (2009).
\bibitem{efg} D.~Ebert, R.~N.~Faustov and V.~O.~Galkin,
Phys.~Rev.~D {\bf 67}, 014027 (2003).

\bibitem{fg} R.~N.~Faustov and V.~O.~Galkin, Z.~Phys.~C {\bf 66}, 119
 (1995); D.~Ebert, R.~N.~Faustov and V.~O.~Galkin,
 Phys.\ Rev.\ D {\bf 73}, 094002 (2006).
\bibitem{bvb} A. M. Badalian, A. I. Veselov and B. L. G. Bakker, {
    Phys. Rev. D} {\bf 70}, 016007 (2004); Yu.~A.~Simonov, {
    Phys. Atom. Nucl.} {\bf 58}, 107 (1995).
\bibitem{lregge}
  D.~Ebert, R.~N.~Faustov and V.~O.~Galkin,
  Eur.\ Phys.\ J.\  C {\bf 47}, 745 (2006);
  Phys.\ Rev.\  D {\bf 79}, 114029 (2009).

\bibitem{shirkov}
D.~Shirkov,
  arXiv:0807.1404 [hep-ph];
  D.~V.~Shirkov and I.~L.~Solovtsov,
  Phys.\ Rev.\ Lett.\  {\bf 79}, 1209 (1997).

\bibitem{hbar} D.~Ebert, R.~N.~Faustov and V.~O.~Galkin,
  Phys.\ Rev.\  D {\bf 72}, 034026 (2005); Phys.\ Lett.\  B {\bf 659},
  612 (2008).




\end{thebibliography}
\end{document}